\documentclass[12pt]{article}
\usepackage{graphics}
     \let\th=\theta  \let\l=\lambda
                 \let\r=\rho
\let\s=\sigma \let\t=\tau   \let\f=\phi 
   \let\o=\omega

\let\O=\Omega  \let\del=\nabla

\def\\{\hfill\break} \let\==\equiv

\let\0=\noindent

\def\ie{{i.e.}}
\let\dpr=\partial

\def\qed{\hfill\raise1pt\hbox{\vrule height5pt width5pt depth0pt}}

\def\be{\begin{equation}}
\def\ee{\end{equation}}
\def\bea{\begin{eqnarray}}\def\eea{\end{eqnarray}}

\begin{document}
\markright{Lorentz...}

\title{The Lorentz Force and the Radiation Pressure of Light}

\author {Tony Rothman{\small\it\thanks{trothman@princeton.edu}}
~and Stephen Boughn{\small\it\thanks{sboughn@haverford.edu}}
\\[2mm]
~ \it Princeton University, Princeton NJ 08544\\
~ \it Haverford College, Haverford PA 19041}

\date{{\small   \LaTeX-ed \today}}

\maketitle

\begin{abstract}

In order to make plausible the idea that light exerts a pressure on matter, some introductory
physics texts consider the force exerted by an electromagnetic wave on an electron.  The argument as
presented is both mathematically incorrect and has several serious conceptual difficulties without obvious resolution at the classical, yet alone introductory, level.  We discuss these difficulties and propose an alternative demonstration.

 \vspace*{5mm} \noindent PACS: 1.40.Gm, 1.55.+b, 03.50De

\end{abstract}
\section{The Freshman Argument}
\setcounter{equation}{0}\label{sec1}
\baselineskip 8mm

 The interaction of light and matter plays a central role, not only in physics itself, but in any freshman
 electricity and magnetism course.  To develop this topic most courses introduce the Lorentz force law, which gives the electromagnetic force acting on a charged particle, and later devote some discussion to Maxwell's equations.  Students are then persuaded that Maxwell's equations admit wave solutions which travel at the speed of light, thus establishing the connection between light and electromagnetic waves.  At this point we  unequivocally state that electromagnetic waves carry momentum in the direction of propagation via the Poynting flux and that light therefore  exerts a radiation pressure on matter.   This is not controversial: Maxwell himself in his {\it Treatise on Electricity and Magnetism}\cite{Maxwell73} recognized that light should
manifest a radiation pressure, but his demonstration is not immediately transparent to modern readers.

At least two contemporary texts, the Berkeley Physics Course\cite{Berk65} and Tipler and Mosca's {\it Physics for Scientists and Engineers}\cite{Tipler}, attempt to make the assertion that light carries momentum more plausible by explicitly calculating  the Lorentz force exerted by an electromagnetic wave on an electron.  In doing so the authors claim---with differing degrees of rigor---to show that a light wave indeed exerts an average force on the electron in the direction of propagation.  Tipler and Mosca, for example, are then able to derive an expression for the radiation pressure  produced by a light wave.

A cursory look at the ``freshman argument," however, which many instructors also present to their classes, shows that in several obvious respects it is simply incorrect and that in other respects it leads rapidly into deep waters.  Nevertheless, one can more  plausibly demonstrate that light exerts a radiation pressure and calculate it in a way that should be accessible to first-year students. It is the purpose of this note to discuss these matters.\\

\begin{figure}[htb] \vbox{\hfil\scalebox{.7}
{\includegraphics{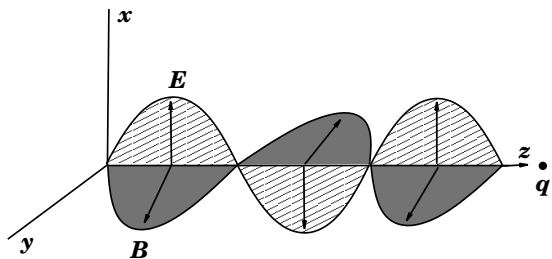}}\hfil} {\caption{\footnotesize{
An electromagnetic wave traveling in the z-direction strikes a point particle with charge $q$.  The $E$-field is taken in the $x$-direction and the $B$-field is taken in the $y$-direction.
 \label{lorentz}}}}
\end{figure}

  Consider, then, the situation shown in figure 1.  We assume that a light wave propagates in the $+z$-direction, that its $E$-field oscillates in the $x$-direction and that its $B$-field oscillates in the $y$-direction.  The wave impinges on a stationary particle with charge $q$, exerting on it a force according to the Lorentz force law.  In units with $c = 1$, the Lorentz force  is
\be
{\bf F} = q(\bf E +  v \times B),\label{lorentz}\\
\ee
 which becomes
\be
{\bf F} = q(E_x  -v_zB_y ){\bf\hat i} + qv_xB_y {\bf\hat k}.\label{lorentz2}
\ee

The ``freshman argument" goes like this:  Assume that $E\sim  sin(\o t)$ and $B  \sim  sin(\o t)$.   The particle is initially accelerated by the $E$-field in the $+x$-direction and acquires a velocity $v_x > 0$.  The magnetic field then exerts a force on the charge equal to $q\bf v \times B$, which points in the $+z$-direction, the direction of propagation of the wave.   The electromagnetic wave therefore carries a momentum in this direction.  The Berkeley Physics Course in fact states, ``...the motion of the charge is mainly due to {\bf E}.  Thus {\bf v} is along {\bf E} and reverses direction at the same rate that {\bf E} reverses direction.  But {\bf B} reverses whenever {\bf E} reverses.  Thus $\bf v \times B$ always has the same sign."\footnote{\cite{Berk65}, vol. 3, p. 362.}

 A moment's reflection, however, shows that the last assertion is simply false. After one-half cycle, both  $\bf E$ and $\bf B$ change sign.  But because during this time the $E$-field has accelerated  the charge entirely in the $+x$-direction the electron at that point still has a positive $x$-velocity. (In other words, the velocity and acceleration are $90^\circ$ out of phase, as in a harmonic oscillator.)  A similar argument holds for the $z$-velocity.  Thus the cross product $\bf v \times B$  reverses sign and now points in the {\it negative} $z$-direction, opposite the direction of propagation.  Furthermore, because there is  an $x$-component to the force, one needs to argue that on average it is zero.

  The Berkeley  authors indeed claim that the first two terms in Eq. (\ref{lorentz2}) average to zero, the first because {\bf E} varies sinusoidally, the second because {\bf B} varies sinusoidally as well and because one ``can assume that the increment of velocity along $z$ during one cycle is negligible, \ie, we can take the slowly increasing velocity $v_z$ to be constant during one cycle."\footnote{ibid.}  With these assumptions the Berkeley authors conclude that the average force on the charge is $\langle F \rangle = q\langle v_x B_y \rangle {\bf\hat k}$.  Although at first glance the result may seem plausible, it is also incorrect because the velocity and magnetic field are  functionally orthogonal  and consequently the time average of their product vanishes.   That this is so, as well as the previous claim, can be seen by a proper integration of the equation of motion (\ref{lorentz2}).   \\

\section{Equation of Motion}
\setcounter{equation}{0}\label{sec2}

To determine the momentum of the charge, which we take to be an electron,  assume the electric and magnetic fields of the light wave are given by ${\bf E} = E_o sin(\o t + \phi){\bf\hat i}$ and ${\bf B }= B_o sin(\o t + \phi){\bf\hat j}$, where $\phi$ is an arbitrary phase angle.  In our units  $E_o = B_o$.   Setting ${\bf F_{lorentz}}= m d{\bf v}/dt$ in Eq. (\ref{lorentz2}) then gives  a pair of coupled ordinary linear first-order  equations for the electron velocity:
\bea
\frac{dv_z}{dt} &=& \o_c \sin(\o t + \phi)v_x\nonumber\\
\frac{dv_x}{dt} &=& \o_c \sin(\o t + \phi)[1-v_z],
\eea
where we have let $qB_o/m \equiv \o_c$, the cyclotron frequency.

These equations have the somewhat surprising analytic solutions
\bea
v_z(t) &=& c_1\cos\left[\frac{\o_c}{\o}\cos(\o t + \phi)\right] + c_2\sin\left[\frac{\o_c}{\o}\cos(\o t + \phi)\right] + 1, \nonumber\\
v_x(t) &=& c_1\sin\left[\frac{\o_c}{\o}\cos(\o t + \phi)\right] - c_2\cos\left[\frac{\o_c}{\o}\cos(\o t + \phi)\right],
\eea
where $c_1$ and $c_2$ are the integration constants.

If we take $v_z(0) = v_x(0) = 0$, which is reasonable and of sufficient generality for our purposes, we find
\[
c_1 = -\cos\left[\frac{\o_c}{\o}\cos\phi\right]\ ; \  c_2 = -\sin\left[\frac{\o_c}{\o}\cos\phi\right]
\]
and the full solutions are therefore
\bea
v_z(t) &=& -\cos\left(\frac{\o_c}{\o}\cos\phi\right)\cos\left[\frac{\o_c}{\o}\cos(\o t + \phi)\right] -\sin\left(\frac{\o_c}{\o}\cos\phi\right)\sin\left[\frac{\o_c}{\o}\cos(\o t + \phi)\right] + 1, \nonumber\\
v_x(t) &=&-\cos\left(\frac{\o_c}{\o}\cos\phi\right)\sin\left[\frac{\o_c}{\o}\cos(\o t + \phi)\right] + \sin\left(\frac{\o_c}{\o}\cos\phi\right)\cos\left[\frac{\o_c}{\o}\cos(\o t + \phi)\right].\nonumber\\\label{sol}
\eea

\begin{figure}[htb] \vbox{\hfil\scalebox{.7}
{\includegraphics{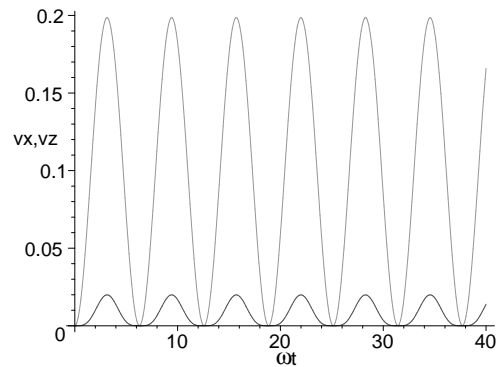}}\hfil} {\caption{\footnotesize{
The $x$-and $z$-velocities vs. $\o t$ for $\o_c/\o = .1$ and $\f = 0$.  Note that from the
small $\o_c/\o$ approximation (see text), $v_x >> v_z$ always; in this case $v_z/v_x = .1$
 \label{lorph0}}}}
\end{figure}

\begin{figure}[htb] \vbox{\hfil\scalebox{.7}
{\includegraphics{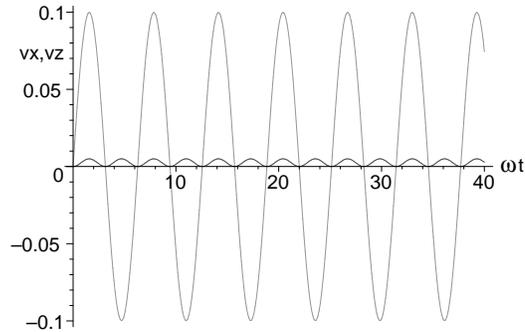}}\hfil} {\caption{\footnotesize{
The same plot as in figure \ref{lorph0} except that $\f = \pi/2$. Note that in this case the
time average of $v_x = 0$.
 \label{lorph2}}}}
\end{figure}

The behavior of these solutions is not exceptionally transparent, but can easily be plotted. Figures 2-4 show several graphs for various values of $\o_c/\o$  and phase angle $\phi$.   Notice that regardless of $\phi$, $v_z$ is always positive, but that there is also a nonzero $v_x$ whose average can be positive, negative or zero depending on $\phi$.  The first case is shown in figure \ref{lorph0} and the third in figure \ref{lorph2}.  Also, in general $v_x >> v_z$.\\

\begin{figure}[htb] \vbox{\hfil\scalebox{.7}
{\includegraphics{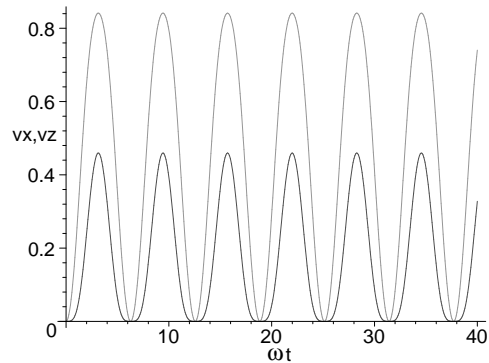}}\hfil} {\caption{\footnotesize{
The same as plot 1 except that $\o_c/\o =.5$
 \label{nrph0}}}}
\end{figure}

 Additional insight into the solutions can be obtained by examining the limit $\o_c/\o < < 1$.  For ordinary light sources at optical frequencies $\o \sim 10^{16}$ rad s$^{-1}$, consideration of the Poynting flux (below) gives
$\o_c/\o \sim  10^{-11}$, and so the limit is well satisfied.  For high-power lasers such as that at the National Ignition Facility, with energy $\sim$ 2 MJ, one can have $\o_c > \o$.  This limit should therefore be avoided. Expanding the solutions (\ref{sol}) to lowest order in $\o_c/\o$ for $\f = 0$ yields
\bea
v_z &\cong& \frac1{2}\left(\frac{\o_c}{\o}\right)^2[\cos(\o t) -1]^2\nonumber\\
v_x &\cong& \left(\frac{\o_c}{\o}\right)[1-\cos(\o t)].\label{wapprox}
\eea

Notice that both $v_z$ and $v_x$ are positive definite, as shown in figure \ref{lorph0}.  Therefore their averages must be as well.  This already contradicts the arguments of \cite{Berk65} that $\langle F_x \rangle = 0$ but that $\langle F_z\rangle\ne  0$.  Notice also that $v_z$ is of order $(\o_c/\o)^2$, while $v_x$ is of order $\o_c/\o$.  This behavior coincides perfectly with the plots, but does suggest that since $v_x^2 \sim (\o_c/\o)^2$, a consistent, relativistic calculation will significantly change $v_z$\cite{Essen03}.  Moreover, the time averages of both  $\dot v_x$ and  $\dot v_z$ vanish to all orders, and so it is in fact impossible to exert a net force on the particle!

One might object to the arguments of this section on the grounds that we have taken  $\bf E$ and $\bf B$  to be simple harmonic $\sim \sin(\o t)$ rather than  wavelike $\sim \sin(kz -\o t)$.  However, it is evident   from Eqs. (\ref{wapprox}) that $kz << \o t $ always and that such corrections are therefore negligible, an assertion borne out by numerical calculations.\\

\section{Interpretation}
\setcounter{equation}{0}\label{sec3}
The question is now whether the behavior just discussed can be reconciled with the classical picture of the Poynting flux.  The Poynting vector in our units is
\be
{\bf S} = \frac{\bf E \times B}{4\pi},
\ee
and the time average is $\langle {\bf S} \rangle = {\mathrm Re} ({\bf E \times B^*})/8\pi$.  $\bf S$  points in the direction of propagation of the electromagnetic wave and in units with $c = 1$ can be regarded interchangeably as power per unit area, energy per unit volume (or pressure) or  momentum flux.   If the freshman argument is correct, then the particle should be accelerated in the direction of the Ponyting vector, but our previous results show clearly that, to the contrary, the particle drifts off in some other direction at a constant average velocity and one searches for a way to explain away this fact.

Unfortunately, there seem to be several deep inconsistencies in the entire approach.  A first is that the freshman derivation is evidently an invalid attempt to apply the standard classical derivation invoked to identify the Poynting flux with electromagnetic momentum, a derivation which   breaks down in the limit we have been considering.  That is, advanced texts such as Jackson\cite{Jackson}, typically begin by considering the Lorentz force on a volume of charges:
\be
\frac{d{\bf p}}{dt} = \int_{vol} (\r {\bf E + J \times B}) d^3 x
\ee
The first step is to eliminate the charge density $\r$ in favor of $\bf E$ via via Gauss's law, $\r = 1/4\pi \nabla \cdot {\bf E}$. One also eliminates $\bf J$ in favor of $\bf \del \times B$ via Amp\`ere's law to find

\bea
\frac{d{\bf p}_{mech}}{dt} &+& \frac{d}{dt}\int_{vol} \frac1{4\pi} ({\bf E \times B)}d^3x\nonumber\\
&=& \frac1{4\pi}
\int_{vol}[{\bf E(\del \cdot E) + E\times(\del \times E)+ B(\del \cdot B)-B\times (\del \times B)]}d^3x.\nonumber\\
\eea

 This is a purely formal result, which after the elimination of $\r$ and $\bf J$ relies only on vector identities.  Since the second term on the left is the only one with a time derivative, one tentatively identifies it with the momentum of the field.

However, the crucial difference between the ``graduate" approach and the freshman method is that in
the graduate approach we are considering a continuous charge distribution.  In the limit of a single charge, the  $\r$ in the Lorentz force law becomes the {\it test} charge distribution, whereas the $\r$ in Gauss's law becomes the {\it source} charge distribution and they cannot be equated.  In the present situation there is not only a single test charge but no source charges whatsoever.  Thus the standard derivation simply cannot be be applied.  Indeed, the only volume one has at one's disposal is the volume of the electron itself, which leads quickly into quantum territory.\\

A second difficulty is that the assumption of plane waves with constant amplitude is an assumption of constant energy and momentum.  If the light wave has constant momentum, how can any be transferred to the electron?  There are many instances in physics where we ignore the backreaction of a recoiling particle on the system.  For instance, according to conservation of momentum, a ball should not bounce off a wall, until one realizes that the ball's change in momentum is absorbed by the earth.

Nevertheless, while to hold the amplitude constant in the current calculation might seem a reasonable approximation, to be totally consistent one should take into consideration the fact that the electron is accelerating and consequently emits radiation, and with that radiation momentum. The customary way to do this  in the nonrelativistic limit is via Thomson scattering, but the differential Thomson scattering cross section for a wave polarized in the $x$-direction is
\be
\frac{d\s}{d\O} = \frac1{2}\left(\frac{e^2}{m}\right)^2(\cos^2\th\cos^2\f + sin^2\f),\label{thom}
\ee
 where $\th$ is the angle between the incident and scattered wave.

 The differential scattering cross section is defined as the ratio of the radiated power per unit solid angle to the incident power per unit area.  We see that the Thomson cross section is absolutely symmetric with respect to reflection through the origin and consequently as much momentum is emitted in the forward as backwards direction. It is therefore far from obvious whether this situation can be corrected in the classical limit.  Indeed, only when one goes to a quantum mechanical derivation (Compton scattering) does one see an asymmetry in the scattering cross section.  In our situation, however, $\hbar\o/m_e \sim10^{-5}$, so it would appear that quantum corrections should be unnecessary.

 What one does practically to get the radiation pressure of light in, say, astrophysical calculations is to multiply the time-averaged Poynting flux $\langle \bf S \rangle$ by the total Thomson cross section $\s_T$.  One can see why this works as follows.  A photon scattered off an electron will have a $z$-momentum $p_z = p_o\cos\th$, for initial momentum $p_o$, and  it therefore removes  $(1-\cos\th) p_o$ from the original  momentum component; the electron must gain the same amount. Multiplying the differential Thomson scattering cross section (\ref{thom}) by $(1-\cos\th)$ and integrating over the sphere, gives exactly the total Thomson scattering cross section
 \be
 \s_T = \frac{8\pi}{3}\left(\frac{e^2}{m}\right)^2.
 \ee
Multiplication by the momentum flux of photons will then give the total force on the electron. Because the Poynting flux {\it is} the momentum flux of photons the same numerical result is obtained by multiplying the Thomson cross section by the time-averaged Poynting flux.  This entire argument, however,  relies on the quantum nature of photons. The Thomson cross section is in actuality the nonrelativistic limit of a cross section that must ultimately be derived from QED, and so we see that the freshman plausibility argument leads quickly to a situation which may have no resolution in the realm of classical physics!

The failing of Thomson scattering is due to the fact that no energy is removed from the original beam. A possible classical ``out" to this situation is simply to assert that the energy  radiated by the electron must be that lost by the incoming beam.  Therefore since $E = p$ for a classical wave, by conservation of momentum the electron must acquire a $z$-momentum exactly as in the Compton scattering case above\cite{Bohm}. While this argument may be valid in terms of conservations laws, it gives no mechanism for transferring the energy from the incident wave to the electron.  Unfortunately, modelling the process as  interference between the incident plane wave and the  spherical wave outgoing from the electron fails to result in any transfer of $z$-momentum from the wave to the charge.  To recover  the Compton  result eventually requires including the radiation-reaction force on the electron, to which we now turn, but because this involves  the classical radius of the electron  it  has already gone beyond the realm of classical electromagnetism.\\

The most ``straightforward" approach to deal with failure of the classical approaches is via the Abraham-Lorentz model, which accounts for the energy radiated by the electron, if  in a somewhat {\it ad hoc} manner.  From the Larmor formula the energy radiated by an accelerated electron over a time $T$ is  $\sim 2e^2a^2T/3$.  Equating this to the kinetic energy  lost by the particle $\sim m a^2T^2 $,  one gets a characteristic time to lose all the energy to radiation:
\[
\t = \frac{2 e^2}{3m}.
\]
This timescale is $2/3$ the time for light to cross the classical radius of the electron, $r_c = e^2/m$, and has a value $\t \sim 10^{-23}$ s.   The total force acting on a particle will now be $m{\bf \dot v = F}_{ext} + {\bf F}_{rad}$, where ${\bf F}_{rad}$ is termed the radiation-reaction force.  Conservation of energy considerations led Abraham and Lorentz to propose that ${\bf F}_{rad} = m\t\bf\ddot v$ (see \cite{Jackson} for more details) and consequently the famous formula
\be
m({\bf \dot v}- \t{\bf \ddot v}) = {\bf F}_{ext}.
\ee

With sufficient massaging, this equation can be applied to the present circumstance to get the desired answer, that the force imparted to the electron by an electromagnetic wave is ${\bf F} = \langle {\bf S}\rangle \s_T$.  Eqs. (\ref{lorentz2}) now become
\bea
\dot v_x - \t\ddot v_x &=& \frac{e}{m}(E_x - v_zB_y)\nonumber\\
\dot v_z - \t\ddot v_z &=& \frac{e}{m}v_xB_y. \label{AL}
\eea

In the nonrelativistic regime $v_z << 1$ and we  ignore the second term on the right in the top equation.  For simplicity we also take both $v_x$ and $v_z$ to be of the form $v = v_oe^{-i\o t}$,  which is of course manifestly untrue according to the  results of \S\ref{sec2}. Then $\dot v_x = -i\o v_x$ and $\ddot v_x = -\o^2 v_x$.   The first of Eqs. (\ref{AL})  becomes
\be
-i\o v_x(1+ i\o \t) \cong \frac{e}{m}E_x,
\ee
or with $\o\t << 1$
\be
v_x \cong \frac{ie}{m\o}E_x(1-i\o \t).
\ee

  With the assumption that $\o_c/\o << 1$  and $\o\t << 1$  we can ignore the $\ddot v_z$ term in the second of Eqs. (\ref{AL}).  Then
\be
\dot v_z \cong \frac{ie^2}{m^2\o}E_xB_y(1-i\o\t).
\ee
For simplicity, take $E_x, B_y$ real. Then we want the time average of the real part of this expression, or
\be
\langle F_z \rangle = \langle m \dot v_z \rangle = \frac{e^4}{m^2}\frac{E_oB_o}{3}  = \langle S \rangle \s_T,
\ee
as fervently desired. The earliest paper we have found that proposes this  calculation is by Page\cite{Page} in 1920, although one suspects that Eddington carried  it out earlier.

Clearly there are a few things left to be desired in this derivation, but it does serve to show that the radiation-reaction force is necessary to get the claimed result.  With slightly more work the conclusion can  be put on a firmer footing via a perturbation calculation\cite{Bohm}: Note that Eqs. (\ref{wapprox}) are the zeroeth-order solutions of Eqs. (\ref{AL}), that is, when  $\t = 0$ and $v_z << 1$ is neglected.    Assume $v_x \equiv v_{x0}+ v_{x1}$ and $v_z \equiv (v_{z0} + v_{z1})<< v_x$,  where the subscript $0$ refers to the zeroeth-order solution and the subscript $1$ refers to the perturbation.  With the help of Eqs. (\ref{wapprox})and (\ref{AL}) it is not too difficult to show that
\be
\dot v_{z1} \cong \o_c v_{x1}\sin(\o t) \cong \o_c^2 \t\sin^2(\o t).
\ee
Taking the time average of this expression vindicates the previous result.  We emphasize, however, that the Abraham-Lorentz model includes an explicit statement about the structure of the electron and hence cannot be regarded as entirely classical; the model is in fact a transition to quantum mechanics and quantum field theory.

\section{Alternative Approach}
\setcounter{equation}{0}\label{sec4}

Despite  the many pitfalls revealed by the above methods, there is a superior and convincing demonstration that light  exerts a pressure on matter, one that should be accessible to freshmen who have had a basic exposure to Maxwell's equations.   The  great advantage of the method is that it avoids consideration of the force acting on a point charge and  can therefore be carried out at the purely classical level.  For this reason it should be adopted by introductory textbook authors.  What follows is a  simplified version of a calculation described by Planck in his {\it Theory of Heat Radiation}\cite{Planck}.

\begin{figure}[htb] \vbox{\hfil\scalebox{.7}
{\includegraphics{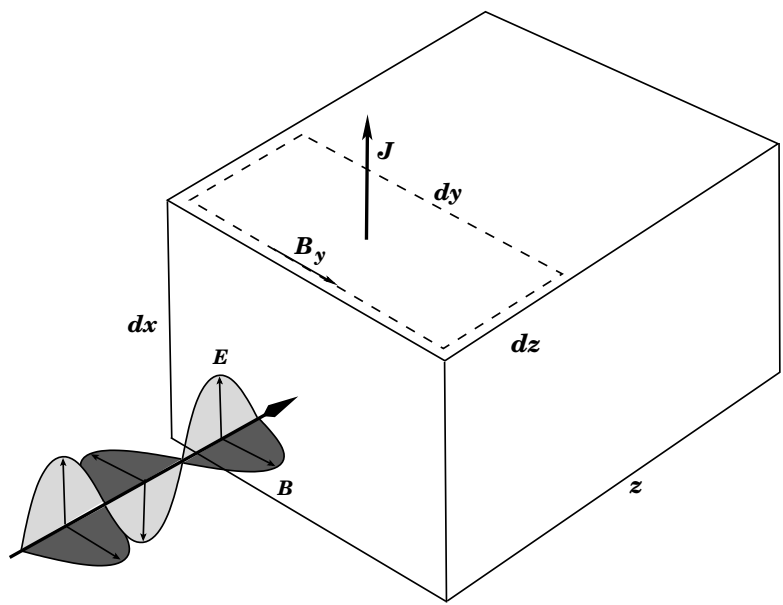}}\hfil} {\caption{\footnotesize{
A light wave traveling in the z-direction strikes an almost perfectly conducting mirror of thickness
$z$, width $dy$ and height $dx$. An Amp\`erian loop in the yz plane is also shown, with the direction of $\bf B$ given by the right-hand rule.
 \label{mirror}}}}
\end{figure}

Consider a light wave propagating, as before, in the $+z$-direction, and which bounces off a mirror, located at $z = 0$ (see figure \ref{mirror}).  We take the mirror to be a near perfect conductor of height $dx$, width $dy$ and thickness $z$.  The electric field of the light will now be a superposition of right- and left-traveling waves:
\be
E_x = E_o\cos(kz -\o t) - E_o\cos(kz + \o t),
\ee
where $k = 2\pi/\l$ is the wave number  and where we have included a phase change on reflection.  (This solution ensures that $E=0$ at the surface of the conductor. Recall that the tangential component of an $E$-field must be continuous across a boundary, and since in the case of a good conductor the interior field essentially vanishes, the exterior field at the boundary must also.)

From the differential form of Faraday's law,\footnote{Most introductory texts use the integral
form of Maxwell's equations.  The derivation can easily be carried out by considering  infinitesimal loops in the xz and yz planes as follows:
The integral form of Faraday's law is $\oint {\bf E \cdot ds} = -d\f/dt$ for magnetic flux $\f$.  In the case of our mirror,
the right-hand rule gives
\[
\oint {\bf E \cdot ds}= E_x(z+dz)dx-E_x(z)dx =
 \frac{dE_x}{dz}dzdx = -\frac{dB}{dt}dxdz,
 \]
  which leads immediately to Eq. (\ref{curlE}).

 Similarly, the integral form of Amp\`ere's law  $\oint {\bf B\cdot ds} = 4\pi I$ leads to Eq. (\ref{Amp}).}
    ${\bf \del \times E}= -{\dpr {\bf B}}/{\dpr t}$  we have

\be
{\bf \del \times E} = \frac{\dpr {E_x}}{\dpr z} {\bf\hat j} = -E_ok[\sin( kz -\o t) -\sin(kz + \o t)]{\bf\hat j}
= -\frac{\dpr {\bf B}}{\dpr t} .\label{curlE}\\
\ee

Integrating with respect to $t$ and remembering that $k \=\o$ in units where $c = 1$ gives
\be
{\bf B} = E_o[\cos(kz -\o t) + \cos(kz + \o t)]{\bf\hat j} = 2B_o\cos(kz)\cos(\o t){\bf\hat j}. \label{B}\\
\ee

Notice that at the boundary, $B = 2B_o\cos(\o t) \ne 0$ and that therefore by Amp\`ere's law, $\oint {\bf B\cdot ds} = 4\pi I$, oscillating currents must be induced near the surface of the mirror.  Since $\bf B$ is in the $\pm y$-direction, the right-hand-rule tells us that these currents will be in the $\pm x$-direction, but that $\bf I \times B$ will always point in the $+z$-direction.  Therefore the Lorentz force due to the light, ${\bf F} = {\bf I} dx\times \bf B$  for a mirror of height $dx$ and total current $I$ will in fact produce a force in the direction of propagation.

We can calculate the magnitude of the force  simply and  plausibly.  The magnitude of the Lorentz force is
$dF = IdxB$, or $dF = JdxdydzB$, for current density $J$.  Now, the differential form of Amp\`ere's law tells us
\be
{\bf \del \times B} = -\frac{\dpr {B_y}}{\dpr z} {\bf\hat i} = 4\pi {\bf J}, \label{Amp}
\ee
or $J = -(1/4\pi) \dpr B_y/\dpr z$.  The Lorentz force therefore becomes
\be
\frac{dF}{dx dy} = -\frac1{4\pi}\frac{\dpr {B_y}}{\dpr z}B_y dz.
\ee
The quantity on the left is of course $dp$, for pressure $p$.  Since the  only spatial dependence of $B$ is on $z$ we can ignore the distinction between the partial and full differentials.  Evidently, since $\dpr B_y/\dpr z$  is connected to $J$, we must interpret $B$ as being the field exerting a force on a given slice within the conductor. Then, if we assume that the magnetic field drops off to zero at infinity, which is certainly true inside a good conductor where the falloff is exponential, the total pressure on the mirror should be
\be
p = -\frac1{4\pi}\int_0^\infty B dB = +\frac1{8\pi} B(0)^2 = \frac1{2\pi}B_o^2 cos^2(\o t),
\ee
where the last equality follows from Eq. (\ref{B}) and the continuity of the tangential component of $\bf B$ across the boundary. The time average of this expression gives
\be
p = \frac{E_oB_o}{4\pi} = 2\langle S \rangle_{incident}
\ee
as desired and where the factor of two is expected due to the recoil of the wave off the mirror.

There are a few tacit assumptions in this derivation that should perhaps be made explicit.  One might wonder, for example, why we use Amp\`ere's law (\ref{Amp}) to calculate the conduction current, rather than Faraday's law $d\f/dt = -\oint {\bf E\cdot ds} = \cal E$, for magnetic flux $\f = B dx dz$ and induced EMF $\cal E$.  Normally,  we would have students use this law to calculate the induced current $I = {\cal E}/R$ in, for example,  a wire loop of resistance $R$. However, in a good conductor $E << B$ and hence $|d\f/dt| = \oint {\bf E\cdot ds} < < {\bf \oint B\cdot ds} = 4\pi I $, the last equality representing Amp\`ere's law.  Furthermore, the $B$-field in Eq. (\ref{Amp}) includes both the incident field and that generated by the induced currents.  It seems unreasonable
that the portion of the $B$-field generated by the induced currents can result in
a net force on the currents themselves (no ``Munchausen effect").  In fact this is the case and a detailed calculation demonstrates that integrated force on the induced currents by the induced $B$-field vanishes.

Nevertheless, with these assumptions the simpler derivation we have presented  appears sound, and it unequivocally  shows that light waves  do exert a pressure on matter in the direction of propagation. \\

In conclusion we might say that, although one does not, and cannot, expect derivations at the freshman level to be
uniformly rigorous, this  case is of particular interest because  the  interaction of light with matter is of fundamental importance.  Moreover, the explanations  presented in textbooks and in the classroom are so seriously flawed that even  students  sometimes notice the difficulties.  Rather than try to paper over these problems with what must be regarded as nonsensical arguments, the occasion would be better exploited to point out that physics is composed of a collection of models that are brought to bear in explaining physical phenomena, but that these models have limited domains of applicability and, as often as not, are inconsistent. \\

{\bf Acknowledgements}

We would like to thank Jim Peebles for suggesting  the mirror argument.

{\small

\end{document}